\documentclass
[showpacs,showkeys,superbib,balancelastpage,prb,10pt,bibnotes,footinbib,preprint,final,a4paper,notitlepage,twocolumn]{revtex4}%
\usepackage{amsfonts}
\usepackage{amsmath}
\usepackage{amssymb}
\usepackage{graphicx}%
\begin{document}
\title[Oscillatory dynamics of a SC VL]{Oscillatory dynamics of a superconductor vortex lattice in high amplitude ac
magnetic fields.}
\author{A.J. Moreno}
\email{amoreno@df.uba.ar}
\affiliation{Laboratorio de Bajas Temperaturas, Departamento de F\'{\i}sica, Facultad de
Ciencias Exactas y Naturales, Universidad de Buenos Aires, Pabell\'{o}n I,
Ciudad Universitaria, C1428EGA Ciudad de Buenos Aires, Argentina.}
\author{S.O. Valenzuela}
\altaffiliation{Present address: Physics Department, Harvard University, Cambridge, MA 02138.}

\affiliation{Laboratorio de Bajas Temperaturas, Departamento de F\'{\i}sica, Facultad de
Ciencias Exactas y Naturales, Universidad de Buenos Aires, Pabell\'{o}n I,
Ciudad Universitaria, C1428EGA Ciudad de Buenos Aires, Argentina.}
\author{G. Pasquini}
\affiliation{Laboratorio de Bajas Temperaturas, Departamento de F\'{\i}sica, Facultad de
Ciencias Exactas y Naturales, Universidad de Buenos Aires, Pabell\'{o}n I,
Ciudad Universitaria, C1428EGA Ciudad de Buenos Aires, Argentina.}
\author{V. Bekeris}
\affiliation{Laboratorio de Bajas Temperaturas, Departamento de F\'{\i}sica, Facultad de
Ciencias Exactas y Naturales, Universidad de Buenos Aires, Pabell\'{o}n I,
Ciudad Universitaria, C1428EGA Ciudad de Buenos Aires, Argentina.}

\begin{abstract}
In this work we study by ac susceptibility measurements the evolution of the
solid vortex lattice mobility under oscillating forces. Previous work had
already shown that in YBCO single crystals, below the melting transition, a
temporarily symmetric magnetic ac field (e.g. sinusoidal, square, triangular)
can heal the vortex lattice (VL) and increase its mobility, but a temporarily
asymmetric one (e.g. sawtooth) of the same amplitude can tear the lattice into
a more pinned disordered state. In this work we present evidence that the
mobility of the VL is reduced for large vortex displacements, in agreement
with predictions of recent simulations. We show that with large symmetric
oscillating fields both an initially ordered or an initially disordered VL
configuration evolve towards a less mobile lattice, supporting the scenario of
plastic flow.

\end{abstract}
\volumeyear{year}
\volumenumber{number}
\issuenumber{number}
\eid{identifier}
\date[Date text]{date}
\received[Received text]{date}

\revised[Revised text]{date}

\accepted[Accepted text]{date}

\published[Published text]{date}

\startpage{1}
\endpage{4}
\maketitle

A rich variety of dc and ac dynamical behaviors arising from the competition
between pinning, thermal and intervortex interactions has been observed in the
vortex lattice (VL) of type II superconductors\cite{blatter}. Forces between
vortices favor the formation of an ordered hexagonal lattice, in opposition to
the disorder that arises from interactions with pinning centers and thermal
forces, leading to defective polycrystalline, or glassy structures.

In driven lattices, external driving forces participate in the formation of
ordered and disordered structures. An example is the experimental
\cite{Kesprb34}and theoretical\cite{KV} evidence of a two step depinning
process of the VL as the driving force is increased. Initially, at low drives,
the lattice undergoes plastic flow, vortices move past their neighbors tearing
the VL and leading to the formation of a disordered lattice, with a high
density of topological defects (e.g. dislocations). Increasing the drive above
a threshold force, F$_{T}$, a dynamical crystallization occurs as proposed by
Koshelev and Vinokur\cite{KV}. At larger forces vortex-vortex interactions
dominate over interactions with pinning centers which are accounted for by an
effective temperature that decreases with the VL velocity.

Memory effects have been observed in low\cite{henderson-andrei}$^{,}%
$\cite{xiaoprl} and high T$_{c}$ materials (HTSC)\cite{sov1}$^{,}$\cite{sov2},
where the resistivity or the apparent critical current density J$_{c}$, are
found to be strongly dependent on the dynamical history of the VL. An increase
in the mobility of the VL when set in motion by a temporarily symmetric (e.g.
sinusoidal) ac field (or current)\cite{henderson-andrei}$^{,}$\cite{xiaoprl}%
$^{,}$\cite{sov4}$^{,}$\cite{sov2} is a characteristic which is common to all
of these experiments. A proposed mechanism\cite{mikitik} for this effect in
YBa$_{2}$Cu$_{3}$O$_{7}$\ (YBCO) crystals, is the annealing of bulk magnetic
gradients in a platelet placed in a perpendicular dc magnetic field by a weak
planar ac magnetic field. A second invoked mechanism is an equilibration
process assisted by the ac magnetic field\cite{ling}$^{,}$\cite{chaddah}.

Transport and ac susceptibility experiments\cite{henderson-andrei}$^{,}%
$\cite{ling}$^{,}$\cite{sov2}$^{,}$\cite{Paltiel} suggest that the response of
a steady driven VL may differ qualitatively from the ac response observed in
measurements involving comparable driving forces. In particular markedly
different effects for temporarily symmetric or temporarily asymmetric ac
drives have been reported\cite{sov2}. It was shown that the application of a
temporarily asymmetric ac magnetic field (e.g. sawtooth) reduces the mobility
of the VL, in contrast to the effect of a symmetric ac field (e.g. square,
sinusoidal, triangular) of similar amplitude and frequency. These effects were
observed to be weakly frequency dependent and to depend strongly on the number
of shaking\ cycles. This is a main result that ruled out an equilibration
process as a possible explanation to the change in vortex mobility\cite{sov2}.
At the same time, the dynamic crystallization scenario becomes inadequate to
describe the observed changes in VL mobility of the solid vortex in twinned
YBCO crystals.

Oscillatory dynamics has been described recently in molecular dynamics
calculations\cite{sov4}, where an oscillatory driving force below the
crystallization threshold is able to order the VL after a number of cycles.
The reordering of the VL is more efficient when vortex excursions are of the
order of the lattice parameter a$_{0}$. This VL shaking\ promotes repeated
interactions between neighbors that heal lattice defects (i.e. the number of
vortices with 5 or 7 first neighbors). At the same time, the average mobility
of the VL increases logarithmically with the number of cycles. An important
result is that for a tiny asymmetry in the amplitude of the force, the vortex
lattice quickly disorders increasing the number of topological defects and the
mobility is rapidly reduced after a few cycles as observed in the
experiments\cite{sov2}. On the other hand, when the period of a symmetrical
force was increased so that the excursion of vortices greatly exceeded the
lattice constant, the VL did not reorder. It was argued that if the period of
the oscillation was large enough, the system should behave as when driven by
steady forces below the threshold force.

In this paper, we investigate the effect of shaking the vortex lattice with
sinusoidal magnetic ac fields starting from a well defined and reproducible
state. In these experiments the excursion amplitude of the vortices is
controlled by the amplitude of the ac magnetic field which is varied between 1
and 150 Oe. We found that for amplitudes below a certain threshold, 20 Oe, the
VL mobility increases as a function of the number of cycles of the shaking
field. However, such an increase in the mobility is not observed above 20 Oe.
Following previous experimental and theoretical results, we interpret this as
an indication that plastic motion dominates the dynamical behavior of the VL.
The reduced mobility would be a consequence of large vortex displacements
which produce vortex lattice tearing and generation of defects.

\qquad We measured the response of the vortex lattice to the shaking field by
means of ac susceptibility measurements. Global ac susceptibility measurements
($\chi_{ac}$ = $\chi$%
%TCIMACRO{\U{b4} }%
%BeginExpansion
\'{}
%EndExpansion
+ j $\chi$\textquotedblright) were made with the usual mutual inductance
technique in two twinned YBa$_{2}$Cu$_{3}$O$_{7}$ single
crystals\cite{muestras}, with typical dimensions 0.6 x 0.6 x 0.02 mm$^{3}$,
and T$_{c}$ $\sim$ 92 K at zero dc field and $\Delta$T$_{c}$ $\sim$ 0.3 K
(10\%-90\% criteria).We have obtained similar results for both crystals but we
show the results obtained for one of them. The ac field, H$_{ac}$, was applied
parallel to the \textbf{c} crystallographic axis and the dc field, H$_{dc}%
,$was applied at $\Theta$=20$%
%TCIMACRO{\U{b0}}%
%BeginExpansion
{{}^\circ}%
%EndExpansion
$ away from the twin boundaries to avoid the effects of correlated
defects\cite{sov1}. In the temperature and field region of interest, high
dissipation ($\chi$") [or low screening ($\chi$')] implies high mobility, and
low dissipation [or high screening] implies low mobility\cite{sov2}.

The experiments to investigate the effect of the amplitude of the
shaking\ field followed the protocol that is described immediately below.
First, the VL lattice was prepared applying the ac configuration field
(H$_{cf}$) for a number of cycles (N $\sim$10$^{5}$, \textit{f} = 10 kHz). A
temporarily asymmetric (sawtooth) H$_{cf}$ was used to prepare a low mobility
configuration, LMC. A temporarily symmetric (sinusoidal) H$_{cf}$ was used to
prepare an initial vortex configuration with high mobility, HMC. It is worth
to note that the lattice was always prepared in the LMC before applying the
symmetric H$_{cf}$ to attain the HMC. The configuration field and the
measuring H$_{ac}$ field were provided by the susceptometer primary coil.
After setting the desired starting configuration, H$_{cf}$ was turned off and
a temporarily symmetric shaking ac field (H$_{sh}$) with an amplitude that
could be varied between 1 Oe and 150 Oe (\textit{f }= 1 Hz) was applied. This
field was supplied by the same electromagnet that provided the dc field. After
shaking the vortex lattice for a number of cycles (N$_{sh}$), H$_{sh}$ was
turned off and the first harmonic of the magnetic ac susceptibility was
measured with a sinusoidal ac field with an amplitude H$_{ac}$ = 2 Oe and
\textit{f} = 10.22 kHz. The measured susceptibility is related to the mobility
of the VL and quantifies the effect of the shaking field in it. With this
protocol we can study the effect of N$_{sh}$ cycles of the shaking field to an
initial vortex configuration with either low (LMC) or high (HMC) mobility.
This procedure is repeated for each measured value of N$_{sh}$.

Figure 1 shows ac susceptibility measurements $\chi$" vs T cooled in dc
(H$_{dc}$ = 2 kOe, $\Theta$ = 20$%
%TCIMACRO{\U{b0}}%
%BeginExpansion
{{}^\circ}%
%EndExpansion
$) and ac fields (H$_{ac}$ = 2 Oe, \textit{f} = 10.22 kHz ). Differences
between VLs with high and low mobility are measured for temperatures below the
melting line (solid VL). We choose to make our measurements at T$\sim$85 K
where a larger difference between the measured signal of the high and low
mobility states is observed.
%TCIMACRO{\FRAME{fhFU}{8.6196cm}{6.0649cm}{0pt}{\Qcb{Ac susceptibility
%measurements $\chi$" vs T. This measurement was made lowering temperature with
%H$_{ac}$ and H$_{dc}$ turned on.}}{\Qlb{fig1}}{fig1.jpg}%
%{\special{ language "Scientific Word";  type "GRAPHIC";
%maintain-aspect-ratio TRUE;  display "ICON";  valid_file "F";
%width 8.6196cm;  height 6.0649cm;  depth 0pt;  original-width 3.3468in;
%original-height 2.3471in;  cropleft "0";  croptop "1";  cropright "1";
%cropbottom "0";
%filename '../usuarios/alex/works/Papier/prbfigs/figs-en-jpg/fig1.JPG';file-properties "XNPEU";}%
%}}%
%BeginExpansion
\begin{figure}
[h]
\begin{center}
\includegraphics[
natheight=2.347100in,
natwidth=3.346800in,
height=6.0649cm,
width=8.6196cm
]%
{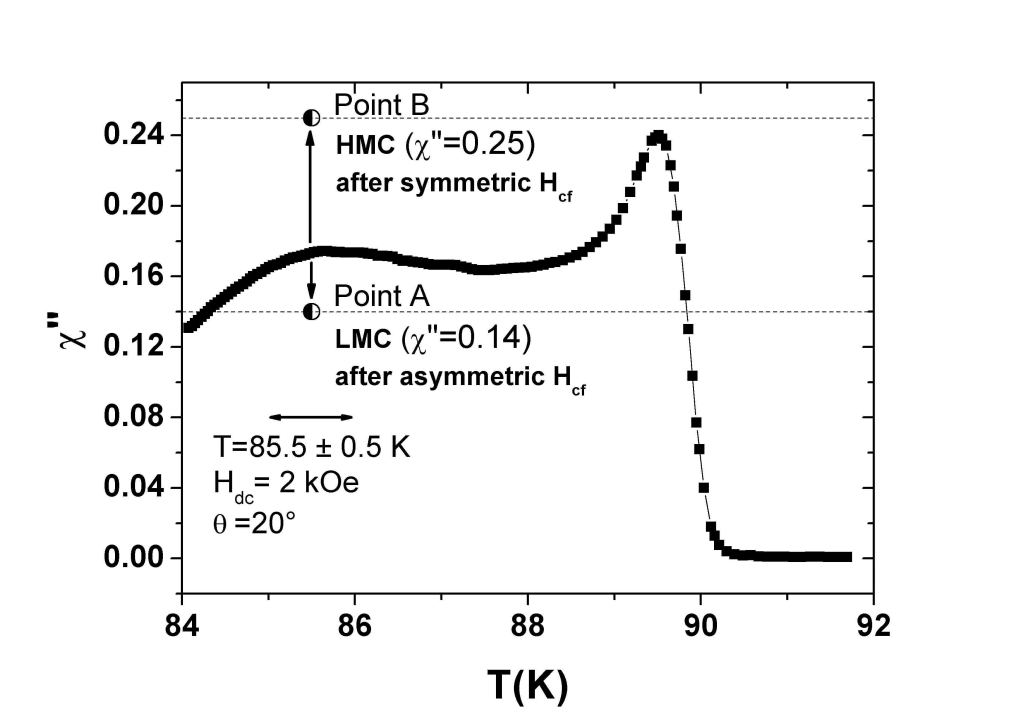}%
\caption{Ac susceptibility measurements $\chi$" vs T. This measurement was
made lowering temperature with H$_{ac}$ and H$_{dc}$ turned on.}%
\label{fig1}%
\end{center}
\end{figure}
%EndExpansion
The lower point (point A) was obtained after field cooling to 85.5 K, turning
off the measuring field and applying 10$^{5}$ cycles of a sawtooth ac magnetic
field, H$_{cf}$ = 7.5 Oe and \textit{f} = 10 kHz. The 7.5 Oe ac field (that
penetrates the sample completely\cite{sov2}) was turned off and ac
susceptibility was measured. The application of the temporarily asymmetric
drive reduced vortex mobility, and we call this a low mobility configuration,
LMC. The higher point (point B) was the dissipation level obtained after
setting a LMC (point A) and then applying 10$^{5}$ cycles of a sinusoidal ac
field, H$_{cf}$ = 7.5 Oe and \textit{f} = 10 kHz. The 7.5 Oe ac field, was
turned off and the ac susceptibility was measured. The mobility of the VL is
clearly enhanced as a result of the application of the sinusoidal field. We
call this vortex state a high mobility configuration, HMC.

Our next experiments were performed at fixed temperature and dc magnetic
field. As anticipated above, the temperature was chosen to correspond with the
low temperature maximum in $\chi$" (see Fig. 1) in order to obtain a large
observable difference between the measured susceptibility for low and high
mobility VL configurations. For our samples, for a dc magnetic field of 2 kOe
and with our selected measuring ac field, this temperature is around T= 85.0
K.
%TCIMACRO{\FRAME{fhFU}{8.6196cm}{6.0649cm}{0pt}{\Qcb{$\chi$" vs number of
%cycles of the shaking field (N$_{sh}$) for different amplitudes, starting from
%a low mobility configuration., LMC.}}{\Qlb{fig2}}{fig2.jpg}%
%{\special{ language "Scientific Word";  type "GRAPHIC";
%maintain-aspect-ratio TRUE;  display "ICON";  valid_file "F";
%width 8.6196cm;  height 6.0649cm;  depth 0pt;  original-width 3.3468in;
%original-height 2.3471in;  cropleft "0";  croptop "1";  cropright "1";
%cropbottom "0";
%filename '../usuarios/alex/works/Papier/prbfigs/figs-en-jpg/fig2.JPG';file-properties "XNPEU";}%
%}}%
%BeginExpansion
\begin{figure}
[h]
\begin{center}
\includegraphics[
natheight=2.347100in,
natwidth=3.346800in,
height=6.0649cm,
width=8.6196cm
]%
{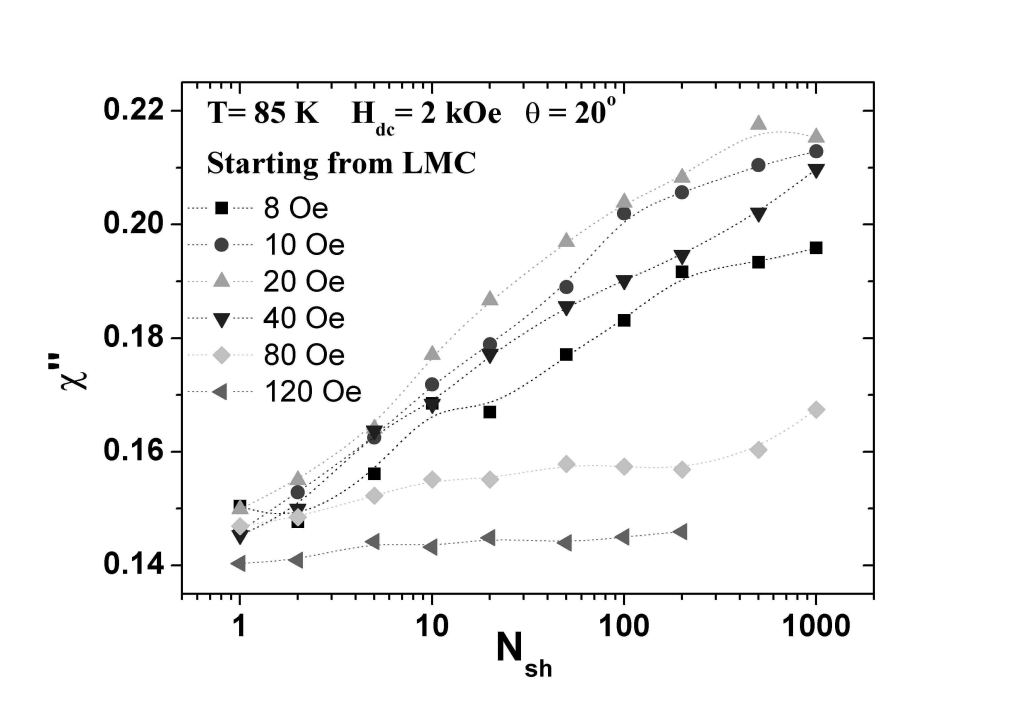}%
\caption{$\chi$" vs number of cycles of the shaking field (N$_{sh}$) for
different amplitudes, starting from a low mobility configuration., LMC.}%
\label{fig2}%
\end{center}
\end{figure}
%EndExpansion

In Fig. 2 we show $\chi$" vs the number of cycles (N$_{sh}$) of the shaking
field. The shaking field was chosen to have different amplitudes (from 8 to a
120 Oe) and was applied to a LMC prepared with H$_{cf}$ =7.5 Oe. Equivalent
results are obtained for $\chi$'. For shaking fields with the lowest amplitude
(8 Oe), $\chi$" increases roughly as the logarithm of N$_{sh}$. The same
dependence is observed for intermediate amplitudes (10-20 Oe) for which it is
also observed that $\chi$" increases with the amplitude of the shaking field
(for a given number of cycles, N$_{sh})$. It is interesting to note that this
increasing trend in $\chi$" does not continue if the amplitude of the shaking
field is further increased. Instead, we observed that $\chi$" reaches a
maximum at around 20 Oe and starts decreasing for larger amplitudes. For
H$_{sh}$ above 80 Oe, the $\chi$" of the final state (after 1000 cycles) is in
fact comparable to the $\chi$" of the initial LMC.

As a larger $\chi$" implies a larger VL mobility, these results indicate that
there is an "optimum" amplitude of H$_{sh}$ ($\sim$20 Oe) for which a maximum
mobility in the VL is obtained (for a given number of cycles). They also show
that a high amplitude symmetric ac field is not effective in reordering and
increasing the mobility of an initially disordered VL. By comparison with
numerical simulations, it appears that large vortex displacements produce
plastic tearing of the VL and the overall response becomes equivalent to the
response of a lattice in a low mobility configuration.%

%TCIMACRO{\FRAME{fhFU}{8.6196cm}{6.0649cm}{0pt}{\Qcb{$\chi$" vs number of
%cycles of the shaking field (N$_{sh}$) for different amplitudes, starting from
%a high mobility configuration HMC.}}{\Qlb{fig3}}{fig3.jpg}%
%{\special{ language "Scientific Word";  type "GRAPHIC";
%maintain-aspect-ratio TRUE;  display "ICON";  valid_file "F";
%width 8.6196cm;  height 6.0649cm;  depth 0pt;  original-width 3.3468in;
%original-height 2.3471in;  cropleft "0";  croptop "1";  cropright "1";
%cropbottom "0";
%filename '../usuarios/alex/works/Papier/prbfigs/figs-en-jpg/fig3.JPG';file-properties "XNPEU";}%
%}}%
%BeginExpansion
\begin{figure}
[h]
\begin{center}
\includegraphics[
natheight=2.347100in,
natwidth=3.346800in,
height=6.0649cm,
width=8.6196cm
]%
{../usuarios/alex/works/Papier/prbfigs/figs-en-jpg/fig3.jpg}%
\caption{$\chi$" vs number of cycles of the shaking field (N$_{sh}$) for
different amplitudes, starting from a high mobility configuration HMC.}%
\label{fig3}%
\end{center}
\end{figure}
%EndExpansion

We also studied the effect of the shaking field on an initially ordered VL
(HMC). In Fig. 3 we show $\chi$" vs the number of cycles of the shaking field
(N$_{sh}$) of different amplitudes starting from the HMC. At low amplitude
magnetic fields (%
%TCIMACRO{\TEXTsymbol{<}}%
%BeginExpansion
$<$%
%EndExpansion
10 Oe) $\chi$" (and the VL mobility) stays approximately constant but at
higher amplitudes (20-80 Oe) an overall reduction in $\chi$" is observed. Even
one cycle is enough to significantly alter the VL configuration. As N$_{sh}$
is increased, the value of $\chi$" seems to go through a shallow minimum but
the $\chi$" never recovers to the initial value at N$_{sh}$=0. Shaking fields
with an amplitude larger that 80 Oe strongly reduce $\chi$" to values which
are comparable to the one in an LMC. As discussed above, the results in Fig. 2
show that a large amplitude symmetric magnetic field is not effective in
reordering and increasing the mobility of the VL. Fig. 3 shows that such an
oscillating magnetic field also distorts an initially ordered VL and reduces
its mobility.

The effects of the amplitude of the shaking field H$_{sh}$ are more clearly
observed in Fig. 4 which shows $\chi$" as a function of the amplitude of the
shaking field for a fixed number of cycles(N$_{sh}$=200). We show measurements
that were performed starting with high and low mobility configurations.
Starting with a LMC it can be seen that the ability of the shaking\ field to
improve VL mobility increases up to a maximum (H$_{sh}$ $\sim$10-20 Oe) and
then decreases. For amplitudes H$_{sh}$ $\sim$80 Oe the VL seems to have
reached a configuration just slightly different from the starting LMC, i.e. a
high amplitude sinusoidal field does not remove VL defects.
%TCIMACRO{\FRAME{fhFU}{8.6196cm}{6.0649cm}{0pt}{\Qcb{$\chi$" vs amplitude of
%shaking field for N$_{sh}$=200, from Low and High mobility configurations.}}%
%{}{fig4.jpg}{\special{ language "Scientific Word";  type "GRAPHIC";
%maintain-aspect-ratio TRUE;  display "ICON";  valid_file "F";
%width 8.6196cm;  height 6.0649cm;  depth 0pt;  original-width 3.3468in;
%original-height 2.3471in;  cropleft "0";  croptop "1";  cropright "1";
%cropbottom "0";
%filename '../usuarios/alex/works/Papier/prbfigs/figs-en-jpg/fig4.JPG';file-properties "XNPEU";}%
%}}%
%BeginExpansion
\begin{figure}
[h]
\begin{center}
\includegraphics[
natheight=2.347100in,
natwidth=3.346800in,
height=6.0649cm,
width=8.6196cm
]%
{../usuarios/alex/works/Papier/prbfigs/figs-en-jpg/fig4.jpg}%
\caption{$\chi$" vs amplitude of shaking field for N$_{sh}$=200, from Low and
High mobility configurations.}%
\end{center}
\end{figure}
%EndExpansion
For a HMC starting state, low amplitudes do not modify the dynamics of the VL.
As H$_{sh}$ is increased above 20 Oe, there is a clear reduction in mobility,
and higher shaking amplitudes configure the VL close to the LMC state. In fact
for H$_{sh}$ $\geq$ 40 Oe the final mobility is independent of the starting
configuration. It is interesting to note that this result is independent of
the frequency of the shaking field in the range tested (0.1 Hz
%TCIMACRO{\TEXTsymbol{<} }%
%BeginExpansion
$<$
%EndExpansion
\textit{f}$_{sh}$
%TCIMACRO{\TEXTsymbol{<} }%
%BeginExpansion
$<$
%EndExpansion
3 Hz). Given that the dissipation in the sample is directly related to the
number of cycles per unit of time, this result implies that the observed
effects are not related to local heating.

Following Ref.7, when the average vortex excursion produced by the shaking
field is comparable to the lattice constant, a$_{0}$, the VL mobility
increases as the vortex lattice orders and moves in an increasingly coherent
way (the calculated number of defects in the lattice and its mobility vary as
the logarithm of the number of cycles of the oscillating force). On the
contrary, calculations predict that when the excursion of vortices greatly
exceeds the lattice constant, the VL mobility is reduced as the plastic motion
tend to increase disorder. In order to relate our results with theoretical
predictions, we estimated the average vortex displacement under the action of
the oscillating ac field. The distance that a vortex at the sample boundary
moves when a field perturbation H$_{sh}$ is applied can be roughly estimated by:

\begin{center}%
\[
\left\langle u\right\rangle \approx\frac{1}{2}\frac{H_{sh}}{H_{dc}}\;r
\]

\end{center}

where r is the sample radius, assuming the magnetic induction B$\sim$H$_{dc}$.
For our experimental conditions, r$\sim$0.3 mm and H$_{dc}$ = 2000 Oe, and
considering a triangular lattice (a$_{0}$ $\sim$0.1$\mu$m),
%TCIMACRO{\TEXTsymbol{<}}%
%BeginExpansion
$<$%
%EndExpansion
u%
%TCIMACRO{\TEXTsymbol{>} }%
%BeginExpansion
$>$
%EndExpansion
$\sim$a$_{0}$ occurs at a shaking field amplitude H$_{sh}$ $\sim$2 Oe.
However, in our experiments the maximum dissipation (implying maximum
mobility) occurs for H$_{sh}$ $\sim$10 Oe (see Fig. 4), so that the above
approximations slightly underestimate the shaking field limit. Note that the
exact value at which the maximum in $\chi$" is observed could depend on the
rigidity of the vortex lattice and the density of pinning centers. The more
rigid the lattice the more difficult is to create defects in it. This implies
that the field amplitude estimated above for the position of the peak is a
lower limit and could increase with vortex rigidity.

We find that our results are qualitatively in accordance with numerical
simulations indicating that if vortices are forced to oscillate with
amplitudes larger than the typical VL parameter, plastic motion introduces
topological defects and reduces mobility. For smaller drives, vortices perform
small excursions interacting repeatedly with neighbors and as the lattice
becomes successively more ordered, its mobility increases.

To conclude, in this paper we have shown that temporarily symmetric vortex
oscillations, forced by sinusoidal ac fields, increase the mobility of the VL
in twinned YBCO crystals. However when the amplitude is larger that a certain
threshold, the temporarily symmetric oscillation reduces the mobility. This is
an indication that large vortex displacements may produce vortex lattice
tearing and the mobility is reduced. We have also shown that a healed lattice
with initial high mobility can be torn by the driven symmetric oscillation, if
vortex displacements are much larger than the VL parameter. The ordered VL
reduces its mobility even with just one oscillation of the shaking field.

This work was supported by: UBACYT X71 and CONICET PID 4634

\bigskip\

\bigskip

\end{document}